\documentclass[reprint,prl,superscriptaddress,showpacs,twocolumn]{revtex4}
\usepackage{units}
\usepackage{amsmath}
\usepackage{amssymb}
\usepackage{graphicx}
\usepackage{bm}
\usepackage{microtype,color}

\newcommand{\be}{\begin{equation}}
\newcommand{\ee}{\end{equation}}
\newcommand{\bea}{\begin{eqnarray}}
\newcommand{\eea}{\end{eqnarray}}

\begin{document}

\title{Using geometry to manipulate long-range correlation of light inside disordered media}

\author{Raktim Sarma}
\affiliation{Department of Applied Physics, Yale University, New Haven, CT, 06520, USA}
\author{Alexey Yamilov}
\email{yamilov@mst.edu}
\affiliation{\textls[-20]{Department of Physics, Missouri University of Science and Technology, Rolla, Missouri 65409,USA}}
\author{Pauf Neupane}
\affiliation{\textls[-20]{Department of Physics, Missouri University of Science and Technology, Rolla, Missouri 65409,USA}}
\author{Hui Cao}
\email{hui.cao@yale.edu}
\affiliation{Department of Applied Physics, Yale University, New Haven, CT, 06520, USA}

\date{\today}

\begin{abstract}

We demonstrate experimentally that long-range intensity correlation for light propagating inside random photonic waveguides can be modified by changing the shape of the waveguide. The functional form of spatial correlation is no longer universal in the regime of diffusive transport and becomes shape-dependent due to the non-local nature of wave propagation. The spatial dependence of the correlation may be asymmetric for light incident from opposite ends of the waveguide. This work opens the door to control non-local effects in mesoscopic transport of waves by manipulating the geometry of random systems.

\end{abstract}


\maketitle

The diffusion model has been widely utilized to describe wave propagation in disordered media, e.g., light in biological tissues, ultrasonic waves through cracked metals, and electron wavefunctions in disordered conductors. It, however, ignores the interference of scattered waves, which lead to many prominent phenomena including Anderson localization, universal conductance fluctuations, and enhanced backscattering \cite{Sheng2,Akkermanbook,2009_Lagendijk_PT}. Extensive theoretical and experimental studies in the past three decades have illustrated that mesoscopic transport of both classical and quantum mechanical waves is governed by wave interference effects \cite{1991_Altshuler,Rossum1}.

One important consequence of wave interferences in random media is the correlations in the fluctuations of scattered intensities \cite{Feng3,Pninichapter}. The interferences between waves scattered along independent paths give rise to intensity correlation on the scale of wavelength, one crossing of paths generates long-range correlation beyond the mean free path, and two crossings leads to an infinite-range correlation \cite{Feng1,Feng2}. The non-local correlations have a direct consequence for the coherent control of light transmission through random media via wavefront shaping~\cite{Mosk_review}, which has advanced rapidly in the past few years due to potential applications to deep tissue imaging~\cite{2011_Wang,2011_Yaqoob,2015_Yu_Wavefront_Shaping_Review}. Indeed focusing light to a single speckle simultaneously brightens nearby speckles, and hence reducing the contrast of focusing~\cite{2012_Davy_Eigenchannels}.
However, the long-range correlation also facilitates the enhancement of total transmission~\cite{2014_Popoff_Eigenvalues} with an optimized wavefront despite of the limited degree of input control ~\cite{2013_Stone_Eigenvalues_with_Absorption}. Moreover, the spatial correlation of intensity inside the random medium~\cite{Shapiro2,Genack1,Lisansk1,PRB} affects focusing and energy deposition into the sample~\cite{Genack_OEx}. Therefore, manipulating the non-local correlation can open up a new avenue to controlling waves inside random media.

Typically the magnitude of long-range correlation is small, but it becomes significant in strongly scattering media, especially when localization regime~\cite{2009_Lagendijk_PT} is approached~\cite{Cwilich,Feng1,Feng2,EPL14,GenackFluct,Muskens2}. Experimentally long-range correlation have been observed not only in space, but also in time, frequency, angle, and polarization, but most measurements are performed on transmitted or reflected light, i.e. {\it outside} the random media \cite{Maret1,Sebbah1,Garcia2,Genack1,Genack2,Lagendijk1,Lagendijk2,Chabanov1,Muskens1}. Modifications of the correlations of transmitted light have been realized with two techniques: (i) varying the probe beam spot size on a wide disordered slab~\cite{Shapiro2,Lagendijk1,Muskens2}, and (ii) inserting a constriction, e.g., a pin hole, inside a random medium~\cite{Maret1,Maret2}.
However, the possibility of manipulating long-range correlation {\it inside} the random media has not been explored.
This is at least in part due to experimental challenge of gaining a noninvasive access to the interior of the medium where the light propagates.

We recently fabricated quasi-two-dimensional random waveguides to probe the transport inside from the third dimension \cite{Dz,SarmaAPL,PRB}. This experimental setup has enabled us to monitor directly how the long-range spatial correlation builds up inside the diffusive system~\cite{PRB}. Moreover, by reducing (or increasing) the width of a rectangular waveguide, we were able to enhance (or suppress) the crossing probabilities of scattering paths throughout the system and, therefore, to modify the magnitude of long-range correlation. However, the functional form of correlation remained unchanged, as it is universal for diffusive quasi-one-dimensional waveguides~\cite{Shapiro2,Lisansk1}.

In this Letter, we experimentally demonstrate an effective approach to manipulating the spatial dependence of long-range intensity correlation {\it inside} random media. This is accomplished by fabricating photonic waveguides with cross-section varying along their length. The functional form of the long-range correlation is modified inside waveguides of different shapes because the crossing probability of scattering paths is affected non-uniformly in space. Our approach enables global optimization via system geometry and it is applicable to other types of waves such as acoustic waves and matter waves. Besides the fundamental importance, manipulating the long-range correlation of waves inside random systems is useful for imaging and focusing into multiply scattering media using wavefront shaping~\cite{Mosk_review, Genack_OEx, Arxiv} because it affects such aspects as focusing contrast, degree of control, as well as energy deposition inside the medium. Therefore, our approach can provide an additional degree of freedom for controlling wave transport in scattering media.

To illustrate the effects of waveguide geometry on long-range spatial correlation, we start with a theoretical analysis of two-dimensional (2D) waveguides. The reflecting sidewalls confine the light inside the waveguide where scattering and diffusion take place within ${\bf r}=(y,z)$ plane with $z$ being the axial direction. Light transport in the random waveguide is diffusive, and the non-local intensity correlation are dominated by the long-range correlation $C_2$~\cite{Cwilich,Feng3}. The 2D correlation function $C_2({\bf r}_1;{\bf r}_2)$ between two points ${\bf r}_1=(y_1,z_1)$ and ${\bf r}_2=(y_2,z_2)$ is calculated with the Langevin approach~\cite{1991_Spivak_Langevin_Approach,Pnini1, Lagendijk1, Lisansk1,Kaveh1}, see~\cite{SI} for details.

Let us consider the simplest case of linear tapering, namely, the waveguide width $W(z)$ increases or decreases linearly along the waveguide axis $z$. Figure 1 shows the magnitude of $C_2$, $C_2({\bf r};{\bf r})$, in three waveguides with $W(z)$ being constant (a), linear increasing (b) or linear decreasing (c). The 2D distributions across the waveguide are clearly different in the three cases, revealing the waveguide geometry has a significant impact on the growth of $C_2$. In Fig. 1(d-f), the correlation functions $C_2(z_1; z_2)$ of the cross-section averaged intensity~\cite{SI} further illustrates the difference: in the waveguide of increasing $W(z)$, the correlation function stay nearly constant for most values of $z_1$ and $z_2$, while in the waveguide of decreasing width, the correlation function exhibits more rapid variation over $z_1$ and $z_2$. These results suggest that the range of spatial correlation is increased (or decreased) in the gradually expanding (or contracting) waveguide, as compared to the waveguide of constant width.

\begin{figure}[htbp]
\centering
\includegraphics[width=3.5in]{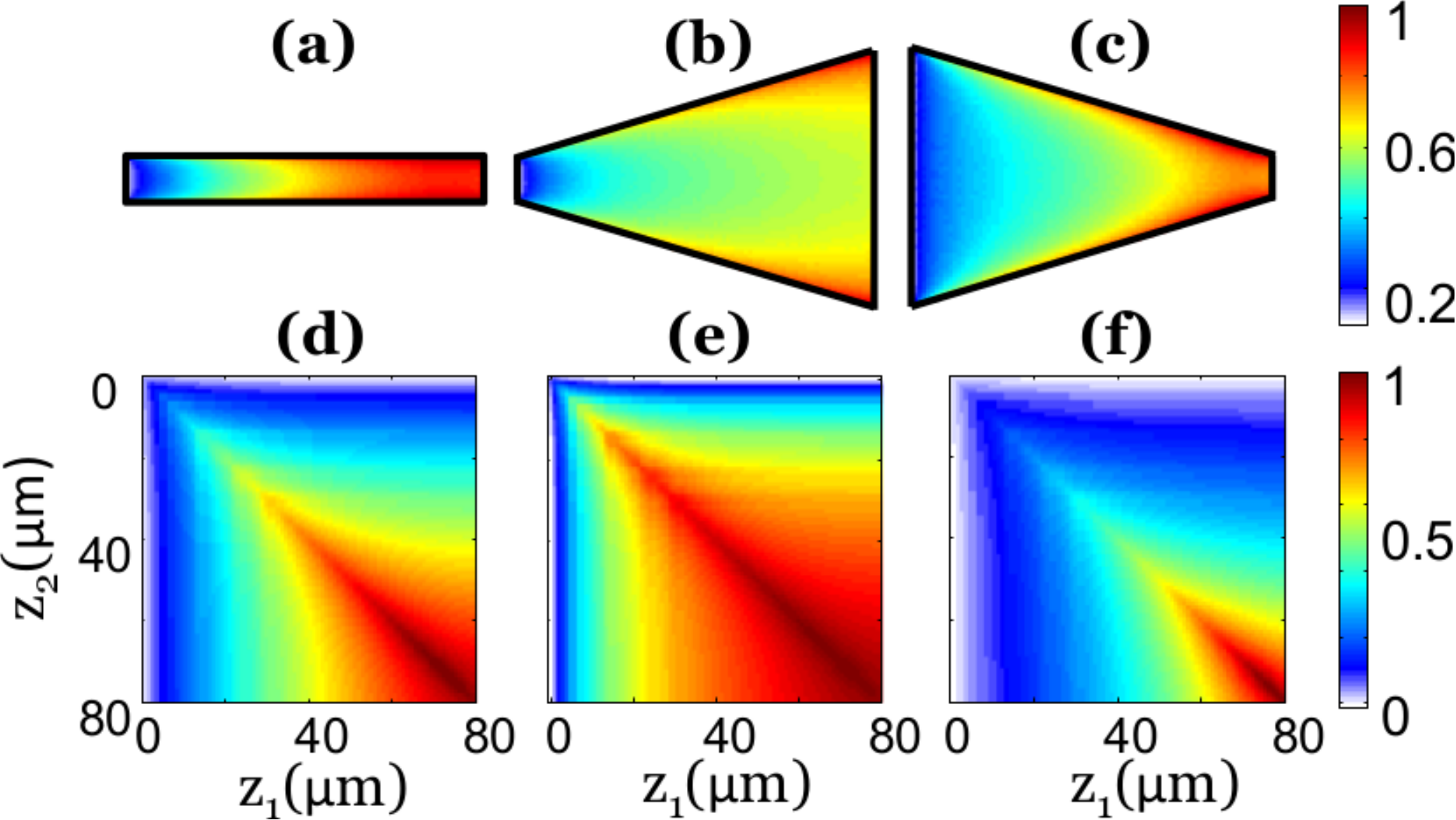}
\caption{(Color online) Calculated spatial long-range intensity correlation for the constant-width and two types of tapered 2D random waveguides. The waveguide length $L$ = 80 $\mu$m, the transport mean free path $\ell$ = 2.2 $\mu$m, and the diffusive absorption length $\xi_a$ = 26 $\mu$m. The waveguide in (a,d) has a constant width $W$ = 10 $\mu$m; in (b,e) $W(z)$ increases linearly from 10 $\mu$m to 60 $\mu$m, while in (c,f) $W(z)$ decreases linearly from 60 $\mu$m to 10 $\mu$m. (a-c) show spatial distribution of the magnitude of long-range correlation function, $C_2({\bf r};{\bf r})$ for three geometries. (d-f) show long-range correlation function $C_2(z_1; z_2)$ of cross-section averaged intensity~\cite{SI} for the same geometries. The maximum value is normalized to 1 for comparison. The differences in these plots reveal that the waveguide geometry has a significant impact on the magnitude and range of $C_2$.
}
\end{figure}

\begin{figure}[htbp]
\centering
\includegraphics[width=3.5in]{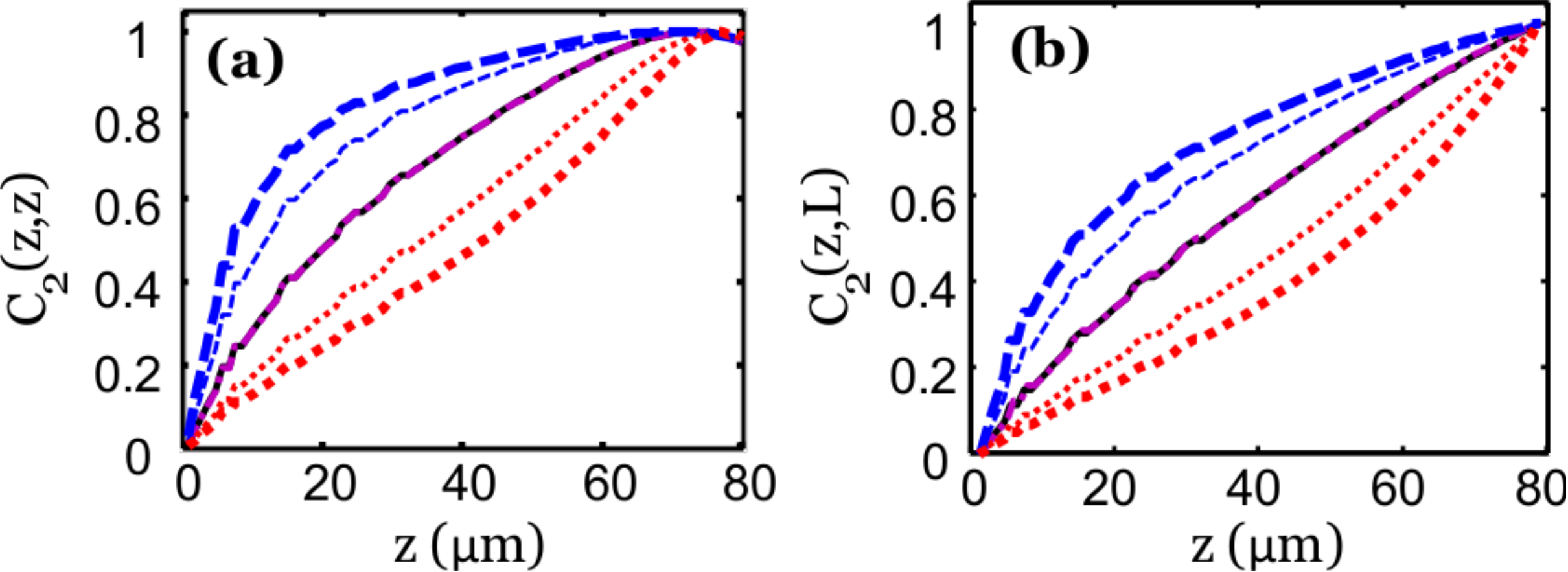}
\caption{(Color online) Comparison of calculated long-range correlation in six waveguides with different degrees of taper:
two with constant widths of 10 $\mu$m (solid black line) and 60 $\mu$m (dash-dotted magenta line);
two with width linearly increasing from 10 $\mu$m (thick dashed blue line) or 20 $\mu$m (thin dashed blue line) to 60 $\mu$m; and
two with width linearly decreasing from 60 $\mu$m to 10 $\mu$m (thick dotted red line) or 20 $\mu$m (thin dotted red line).
Other parameters are the same as in Fig. 1.
Both $C_2(z; z)$ (a) and $C_2(z;L)$ (b) clearly demonstrate that while the functional form of long-range correlation is universal for uniform waveguides, it is strongly modified in the tapered ones.
}
\end{figure}

For a more quantitative comparison, the magnitude of $C_2$ of the cross-section averaged intensity, i.e., $C_2(z;z)$, is plotted in Fig. 2(a) for six waveguides of same length but different geometry. To compare the shape of these curves, the maximum value of each curve is set to 1. After the normalization, the two curves for the constant widths of 10 $\mu$m and 60 $\mu$m coincide and agree to the universal functional form. In the expanding waveguide, $C_2(z; z)$ increases more rapidly at the beginning and levels off when light diffuses deeper into the waveguide. This is attributed to the higher crossing probability of scattering paths near the front end of the waveguide where the cross section is narrower. As the width increases with $z$, the crossing probability is reduced, and $C_2$ rises more slowly. The contracting waveguide exhibits the opposite trend: the magnitude of $C_2$ grows more quickly in the second half of the waveguide due to enhanced crossing probability. The larger the tapering of the waveguide cross section, the bigger the change in the spatial dependence of $C_2$.

Figure 2(b) plots the correlation function $C_2(z;L)$ for two points $z$ and $L$ of cross-section averaged intensity of the six waveguides studied above. After normalizing the maximum value to 1, $C_2(z;L)$ for the constant-width waveguides coincide, whereas in the expanding waveguide, it increases more rapidly with $z$ in the first half of the waveguide, and then increases at a much lower rate in the second half. In the waveguide with decreasing width, the behavior is opposite.

These results confirm that the functional form of long-range correlation is modified by tapering of the waveguide. The deviation from the universal functional form in the constant-width waveguides is larger when the degree of tapering is stronger. The difference in the correlation functions in expanding and contracting waveguides reveals that $C_2({\bf r}_1;{\bf r}_2)$ is no longer symmetric because one waveguide is a mirror image of the other. In other words, the long-range intensity correlation function for light input from one end of the tapered waveguide is different from that with input from the other end. This behavior is distinct from that of the constant-width waveguide whose two ends are equivalent.

Next, we conduct the experiments. The 2D disordered waveguides are fabricated in a silicon-on-insulator (SOI) wafer with a 220 nm silicon layer on top of a 3 $\mu m$ buried oxide. The patterns are written by electron beam lithography and etched in an inductively-coupled-plasma (ICP) reactive-ion-etcher (RIE). Each waveguide contain a 2D random array of air holes that scatter light. The air hole diameters are 100 nm and the average (center-to-center) distance of adjacent holes is 390 nm. The waveguide walls are made of triangle lattice of air holes (the lattice constant of 440 nm, the hole radius of 154 nm) that has a complete 2D photonic bandgap for the in-plane confinement of light.

The monochromatic beam from a tunable CW laser source (HP 8168F) is coupled into the empty waveguide by an objective lens of numerical aperture (NA)  0.4. The light is transverse-electric (TE) polarized, i.e., the electric field is in the plane of the waveguide.
After propagating through the empty waveguide, the light is incident onto the random array of air holes inside the waveguide.
The front end of the random array is uniformly illuminated along the $y$ direction. The light undergoes multiple scattering in the 2D plane of waveguide. Some of the light is scattered out of plane and imaged by a 50$\times$ objective lens (NA = 0.42) onto an InGaAs camera (Xeva 1.7-320).

From the optical image, the spatial distribution of light intensity inside the waveguide $I(y,z)$ is extracted. To smooth out the short-range fluctuations, $I(y,z)$ is averaged over the cross-section of the waveguide to obtain the cross-section-averaged intensity $I_v(z)$. The spatial intensity correlation $C(z_1,z_2)$ is then computed from $I_v(z)$.
After removing the short-range contributions, $C(z_1,z_2)$ is dominated by long-range correlation $C_2$. The contribution of $C_3$, which is on the order of $1/g^2$, is negligible as $g \gg 1$ in our waveguides.

The relevant parameters for light transport in the disordered waveguide are the transport mean free path \textit{l} and the diffusive dissipation length $\xi_a$. The dissipation results from out-of-plane scattering, which can be treated similarly as absorption \cite{Dz}. From the disordered waveguides with constant width, we find $\xi_a = 26$ $\mu m$ and $\ell = 2.2$ $\mu m$ by fitting the measured $I(z)$ and $C(z_1,z_2=z_1)$ \cite{SI}. The tapered waveguides have the same density and diameter of air holes, thus the values of $\xi_a$ and $\ell$ $\mu m$ are identical.

Figure 3(a,b) are the scanning electron microscope (SEM) images of an expanding waveguide and a contracting waveguide. The measured correlation functions for the cross-section averaged intensity inside the two waveguides, $C(z_1=z, z_2 = L)$, are shown in Figure 3(c). To compare the functional form, the measured $C(z,L)$ is normalized. The experimental data clearly show that the dependence of $C(z,L)$ on $z$ is very different for the two tapered waveguides, which agree well to the calculation results.

Since the waveguide geometry in Fig. 3(b) is the mirror image of the one in Fig. 3(a), the $C(z,L)$ for light input from the left end of the former is equivalent to that with input from the right end of the latter. As $C$ is dominated by long-range correlation function, this result implies $C_2$ becomes asymmetric. Therefore, by tapering the waveguides we can modify the crossing probabilities of scattering paths and vary the long-range correlation.

\begin{figure}[htbp]
\centering
\includegraphics[width=3.5in]{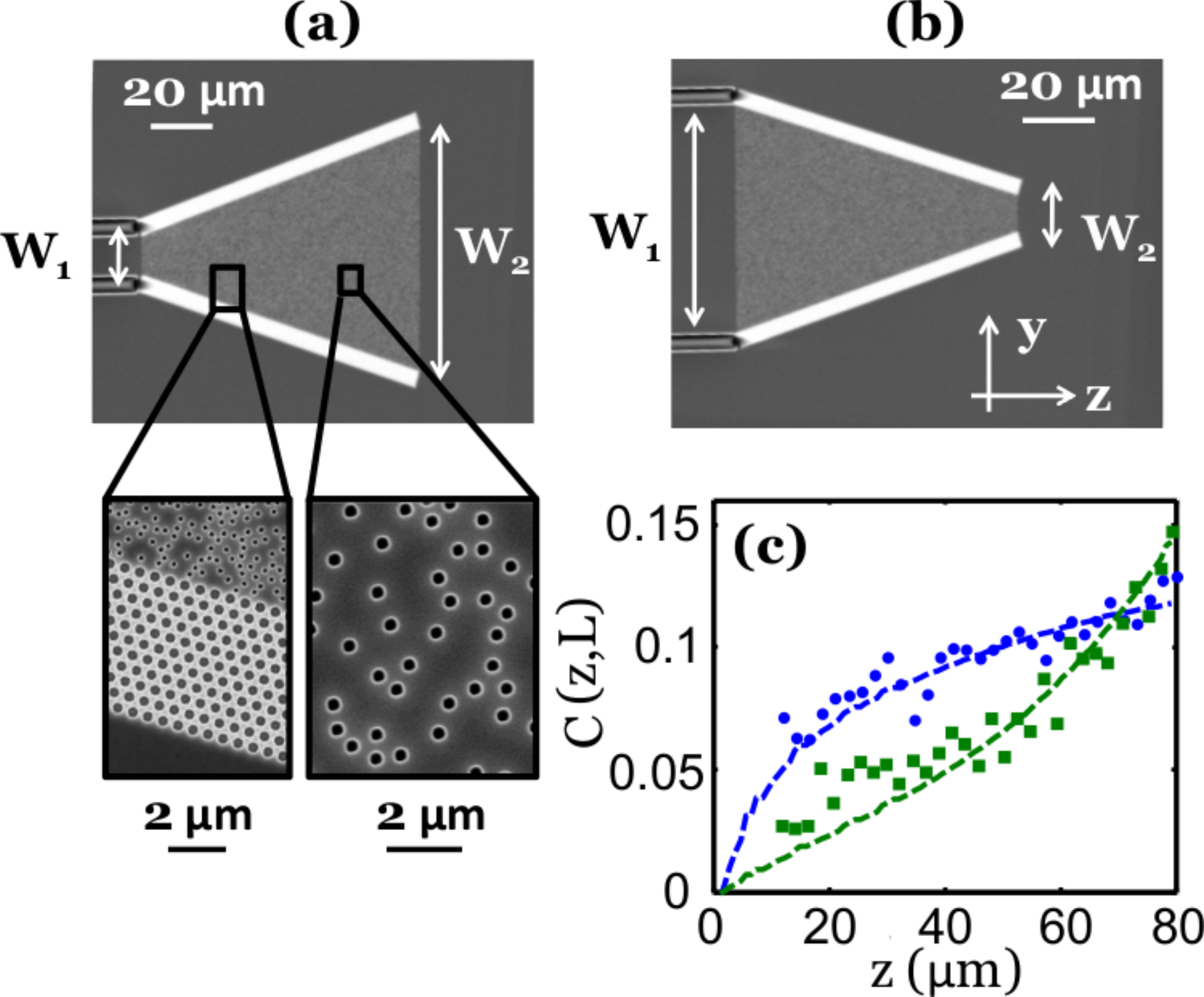}
\caption{Experimental measurement of long-range intensity correlation inside the tapered waveguides. (a,b) Top-view SEM images of fabricated quasi-2D disordered waveguides with linearly increasing (a) or decreasing (b) width. The width of waveguide in (a) increases from 10 $\mu$m to 60 $\mu$m, and in (b) it is opposite. Both have the same length $L$ = 80 $\mu$m. (c) Magnified SEM images showing the air holes distributed randomly in the tapered section of the waveguide and the triangle lattice of air holes in the reflecting sidewalls. (d) Measured long-range correlation function for the cross-section-averaged intensity $C(z, L)$ inside the tapered waveguides shown in (a) and (b). The blue circles (green squares) represent experimental data for the waveguides with increasing (decreasing) width, and the dashed lines are theoretical results. }
\end{figure}

Finally, we modulate the waveguide cross section in a  non-monotonic manner for further manipulation of long-range intensity correlation inside the random waveguide. The waveguide shown in Fig. 4(a) has the width $W$ increasing linearly in the first half of the waveguide and decreasing in the second half. This geometry, unlike the tapered waveguides studied above, is symmetric with respect to the center ($z = L/2$), thus the spatial intensity correlation function is the same for light incident from either end of the waveguide. Figure 4(b) shows the spatial distribution of light intensity inside the waveguide with input from the left end. The short-range intensity fluctuations seen in Fig. 4(b) are smoothed out after averaging the intensity over the cross section, leaving only the long-range contributions to the intensity correlation function $C(z_1, z_2)$. Figure 4(c) plots $C(z,  L)$, which increases initially at a slow rate as $z$ approaches $L/2$, but turns into a sharp rise once $z$ passes $L/2$ and approaches $L$.
This is because the crossing probability of scattering paths is first reduced as the waveguide is expanding in $z<L/2$, and then enhanced in $z>L/2$ as the cross section decreases. Therefore, the crossing probability can be controlled by modulating the waveguide width, which changes the spatial dependence of long-range correlation function. Figure 4(d) shows the intensity correlation function $C(z_1=z, z_2 = L/2)$. It first increases monotonically as $z$ moves from $0$ to $L/2$, and then decreases slightly for $z$ from $L/2$ to $L$. The experimental data (solid circles) are in good agreement to the theoretical results (dashed lines) in Fig. 4(c,d).

\begin{figure}[htbp]
\centering
\includegraphics[width=3.5in]{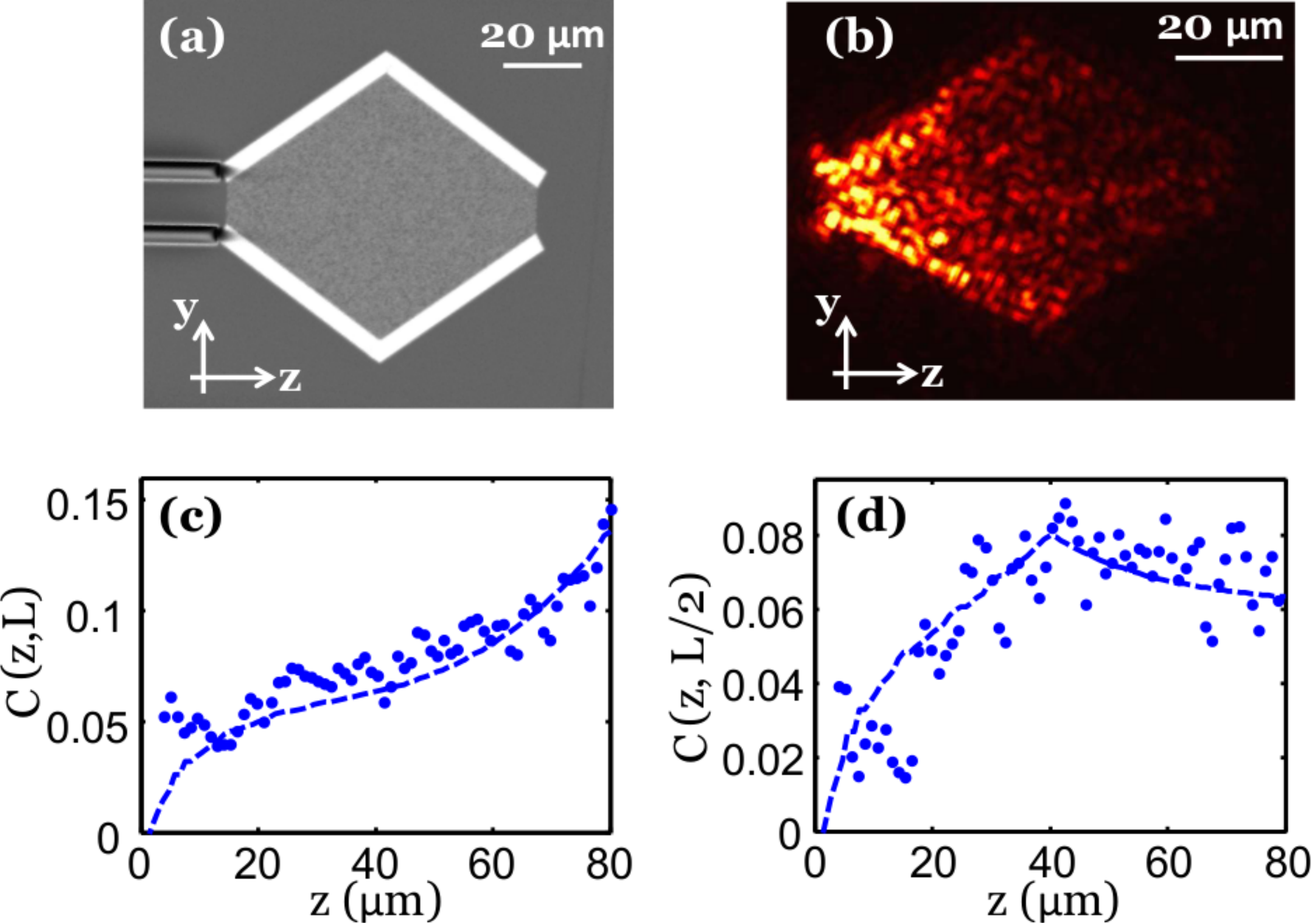}
\caption{Long-range correlation in a quasi-2D disordered waveguide whose width varies non-monotically.
(a) Top-view SEM image showing the waveguide width $W$ increases linearly from 10 $\mu$m at $z$ = 0 to 60 $\mu$m at $z$ = 40 $\mu$m and then reduces linearly down to 10 $\mu$m at $z$ = 80 $\mu$m. Other structural parameters are the same as the waveguides in Fig. 3.
(b) An optical image of the intensity of scattered light from the disordered waveguide. The wavelength of the probe light is 1510 nm. (c) Long-range correlation function $C(z, L)$ for the cross-section averaged intensities at $z$ and $L$ in the waveguide shown in (a). $C(z, L)$ displays a sharp change in the growth rate before and after $z$ passes $L/2$. (d) Long-range correlation function $C(z, L)$ for the cross-section averaged intensities at $z$ and $L/2$ in the waveguide shown in (a). $C(z, L/2)$ increases monotonically in the first half of the waveguide and decreases slightly in the second half. In (c,d), solid circles represent experimental data and the dashed curves are obtained by numerical calculation.
}
\end{figure}

In summary, we have demonstrated an efficient control of long-range intensity correlation inside random media by tailoring the shape of a random system without modifying the structural disorder. By modulating the cross section of a disordered waveguide, we manipulate the crossings of the scattering paths inside the waveguide and change the functional form of long-range correlation from the universal one for constant-width waveguides.
In addition, the long-range correlation function can be made asymmetric with the spatial dependence different for light incident from one end or the other end of the waveguide. This work opens a new direction for coherent control of wave transport in random media.

\begin{acknowledgments}
We thank B.~Shapiro and A.~D.~Stone for stimulating discussions and Michael Rooks for suggestions regarding sample fabrication. This work is supported by the National Science Foundation under Grants No. DMR-1205307 and DMR-1205223. Facilities use is supported by YINQE and NSF MRSEC Grant No. DMR-1119826.

\end{acknowledgments}


\begin{thebibliography}{99}

\bibitem{Sheng2} P. Sheng, {\it Introduction to Wave Scattering, Localization, and Mesoscopic Phenomena} (Academic,Boston, 1995).
\bibitem{Akkermanbook}E. Akkermans and G. Montambaux, {\it Mesoscopic Physics of Electrons and Photons} (Cambridge University Press, Cambridge, 2007).
\bibitem{2009_Lagendijk_PT} A.~Lagendijk and B.~van~Tiggelen and D.~S.~Wiersma, {\it Phys. Today} {\bf 62}, 24 (2009).
\bibitem{1991_Altshuler}B.~L.~Altshuler and P.~A.~Lee and R.~A.~Webb, {\it Mesoscopic Phenomena in Solids} (North Holland, Amsterdam, 1991).
\bibitem{Rossum1} M. C. van Rossum, and T. M. Nieuwenhuizen, {\it Rev. Mod. Phys.} {\bf 71}, 313 (1999).
\bibitem{Feng3} R. Berkovits and S. Feng, {\it Phys. Rep.} {\bf 238}, 135 (1994).
\bibitem{Pninichapter}R. Pnini, {\it Correlation of Speckle in Random Media, Proceedings of the International Physics School on Waves and Imaging through Complex Media}, 391–412 (1999: Cargèse, France) edited by P Sebbah (Kluwer Academic Publishers, Dordrecht, 2001).
\bibitem{Feng1}S. Feng, C. Kane, P. A. Lee, and A. D. Stone, {\it Phys. Rev. Lett.} {\bf 61}, 834 (1988).
\bibitem{Feng2}S. Feng and P. A. Lee, {\it Science} {\bf 251}, 633 (1991).
\bibitem{Mosk_review}A. P. Mosk, A. Lagendijk, G. Lerosey, and M. Fink, {\it Nature Photonics} {\bf 6}, 283 (2012).
\bibitem{2011_Yaqoob}Z.~Yaqoob, D.~Psaltis, M.~S.~Feld, and C.~Yang, {\it Nat. Photon.} {\bf 2}, 110 (2008).
\bibitem{2011_Wang}X.~Xu, H.~Liu, and L.~V.~Wang, {\it Nat. Photon.} {\bf 5}, 154 (2011).
\bibitem{2015_Yu_Wavefront_Shaping_Review}H.~Yu, J.~Park, K.~Lee, J.~Yoon, K.~Kim, S.~Lee, and Y.~K.~Park, {\it Curr. Appl. Phys.} {\bf 15}, 632 (2015).
\bibitem{2012_Davy_Eigenchannels}M.~Davy, Z.~Shi, and A.~Z.~Genack, {\it Phys. Rev. B} {\bf 85}, 035105 (2012).
\bibitem{2014_Popoff_Eigenvalues}S.~M.~Popoff, A.~Goetschy, S.~F.~Liew, A.~D.~Stone, and H.~Cao, {\it Phys. Rev. Lett.} {\bf 112}, 133903 (2014).
\bibitem{2013_Stone_Eigenvalues_with_Absorption}A.~Goetschy, and A.~D.~Stone, {\it Phys. Rev. Lett.} {\bf 111}, 063901 (2013).
\bibitem{Shapiro2}R. Pnini and B.Shapiro, {\it Phys. Rev. B} {\bf 39}, 6986 (1989).
\bibitem{Genack1}A. Z. Genack, N. Garcia, and W. Polkosnik, {\it Phys. Rev. Lett.} {\bf 65}, 2129 (1990).
\bibitem{Lisansk1}A. A. Lisyansky and D. Livdan, {\it Phys. Rev. B} {\bf 47}, 14157 (1993).
\bibitem{PRB} R. Sarma, A. Yamilov, P. Neupane, B. Shapiro, and H. Cao, {\it Phys. Rev. B} {\bf 90}, 014203 (2014).
\bibitem{Genack_OEx} X. Cheng and A. Z. Genack, {\it Optics Letters} {\bf 39}, 6324 (2014).

\bibitem{Cwilich} M.~J.~Stephen and G.~Cwilich, {\it Phys. Rev. Lett.} {\bf 59}, 285 (1987).
\bibitem{GenackFluct}  A.~A.~Chabanov, M.~Stoytchev and A.~Z.~Genack, Nature {\bf 404}, 850 (2000).
\bibitem{Muskens2}T. Strudley, T. Zehender, C. Blejean, E. Bakkers, and O. L. Muskens, {\it Nat. Photonics.} {\bf 7}, 413 (2013).
\bibitem{EPL14}C. P. Lapointe, P. Zakharov, F. Enderli, T. Feurer, S. E. Skipetrov, and F. Scheffold, {\it EPL} {\bf 105}, 34002 (2014).

\bibitem{Lagendijk2}M. P. van Albada, J. F. de Boer, and A. Lagendijk, {\it Phys. Rev. Lett.} {\bf 64}, 2787 (1990).
\bibitem{Lagendijk1}J. F. de Boer, M. P. van Albada, A. Lagendijk, {\it Phys. Rev. B} {\bf 45}, 658 (1992).
\bibitem{Garcia2}N. Garcia, A.Z. Genack, R. Pnini, B. Shapiro, {\it Phys. Lett. A} {\bf 176}, 458 (1993).
\bibitem{Maret1}F. Scheffold, W. Hartl, G. Maret, and E. Matijevic, {\it Phys. Rev. B} {\bf 56}, 10942 (1997).
\bibitem{Genack2}P. Sebbah, R. Pnini, and A. Z. Genack, {\it Phys. Rev. E} {\bf 62}, 7348 (2000).
\bibitem{Sebbah1}P. Sebbah, B. Hu, A. Z. Genack, R. Pnini, and B. Shapiro, {\it Phys. Rev. Lett.} {\bf 88}, 123901 (2002).
\bibitem{Chabanov1}A. A. Chabanov, N. P. Tregoures, B. A. van Tiggelen, and A. Z. Genack, {\it Phys. Rev. Lett.} {\bf 92}, 173901 (2004).
\bibitem{Muskens1}O. L. Muskens, T. van der Beek, and A. Lagendijk, {\it Phys. Rev. B} {\bf 84}, 035106 (2011).
\bibitem{Maret2} F. Scheffold and G. Maret, {\it Phys. Rev. Lett.} {\bf 81}, 5800 (1998).
\bibitem{Dz} A.Yamilov, R. Sarma, B. Redding, B. Payne, H. Noh, and H.Cao  {\it Phys. Rev. Lett.} {\bf 112}, 023904 (2014).
\bibitem{SarmaAPL} R. Sarma, A. Yamilov, T. Golubev, and H. Cao, {\it Appl. Phys. Lett.} {\bf 105}, 041104 (2014).

\bibitem{Arxiv} N. Fayard, A. Caze, R. Pierrat, and R. Carminati, {\it Arxiv:} {\bf 1504.06267} (2015).

\bibitem{1991_Spivak_Langevin_Approach}B.~Z.~Spivak and A.~Y.~Zyuzin, {\it Mesoscopic Phenomena in Solids}, edited by B.~L.~Altshuler, P.~A.~Lee, and R.~A.~Webb, (Elsevier Science Publishers, Amsterdam, 1991) Chap. 2.
\bibitem{Pnini1}R. Pnini and B. Shapiro, {\it Phys. Lett. A} {\bf 157}, 265 (1991).
\bibitem{Kaveh1}E. Kogan and M. Kaveh, {\it Phys. Rev. B} {\bf 45}, 1049 (1992).
\bibitem{SI}Supplemental Material.

\end{thebibliography}
 \end{document}